\documentclass[aps,prd,twocolumn,groupedaddress]{revtex4-1}
\usepackage{hyperref}

\usepackage{amsmath}
\usepackage{amssymb}
\usepackage{longtable}
\usepackage{graphicx}

\def\be{\begin{equation}}
\def\ee{\end{equation}}
\def\bg{\begin{eqnarray}}
\def\en{\end{eqnarray}}

\begin{document}

\title{Variations of nuclear binding with quark masses}
\author{M.~E.~Carrillo-Serrano}
\affiliation{CSSM and ARC Centre of Excellence for Particle Physics at the Tera-scale,\\
School of Chemistry and Physics,
University of Adelaide, Adelaide SA 5005, Australia
}
\author{I.~C.~Clo\"et}
\affiliation{CSSM,
School of Chemistry and Physics,
University of Adelaide, Adelaide SA 5005, Australia
}
\author{A.~W.~Thomas}
\affiliation{CSSM and ARC Centre of Excellence for Particle Physics 
at the Terascale,\\
School of Chemistry and Physics,
University of Adelaide, Adelaide SA 5005, Australia
}
\author{K.~Tsushima}
\affiliation{CSSM,
School of Chemistry and Physics,
University of Adelaide, Adelaide SA 5005, Australia
}
\author{I.~R.~Afnan}
\affiliation{
School of Chemical and Physical Sciences,
Flinders University,
GPO Box 2100, Adelaide SA 5001
Australia
}

\date{\today}

\begin{abstract}
We investigate the variation with light quark mass
of the mass of the nucleon 
as well as the masses of the mesons commonly used in a one-boson-exchange
model of the nucleon-nucleon force. Care is taken to evaluate the meson mass
shifts at the kinematic point relevant to that problem. Using these results,
the corresponding changes in the energy of the $^1$S$_0$ anti-bound state, 
the binding energies of the deuteron, triton and selected finite nuclei 
are evaluated using a one-boson exchange model.
The results are discussed in the context of possible corrections to the
standard scenario for big bang nucleosynthesis in the case where, as
suggested by recent observations of quasar absorption spectra, the
quark masses may have changed over the age of the Universe.

\end{abstract}


\maketitle
\section{Introduction}
In the last decade there has been considerable interest in the
possibility that the fundamental ``constants'' of Nature may actually change
with time~\cite{Uzan:2002vq}. 
Although it remains controversial, there is growing evidence
that the fine structure constant may have varied by an amount of order
a few parts in $10^{-5}$ over a period of 5--10 billion 
years~\cite{Dzuba:1999zz,Webb:1998cq,Murphy:2003hw,Sandvik:2001rv}. It has
even been suggested that this variation may have a dipole structure as
we look back in different directions~\cite{Webb:2010hc}. 
Although this possible variation
is quite small, within the framework of most attempts at grand unification,
a variation of $\alpha$ implies considerably larger percentage changes in
quantities such as $\Lambda_{\rm QCD}$ and in the quark 
masses~\cite{SS,Bekenstein:1982eu,Olive:2002tz}. For
example, in Ref.~\cite{SS} it was shown that 
the variation $\delta m_q/m_q$
would be of order 38 times that of $\delta \alpha/\alpha$.

In the light of these developments it is very natural to ask what other
signatures there may be for such changes. These may, for example, be
the consequent changes in hadron masses or magnetic 
moments~\cite{Flambaum:2002wq,Flambaum:2004tm,Cloet:2008wg}.
Indeed, in some cases the level of precision possible in modern atomic,
molecular and optical physics means that it may even be feasible to
detect the minute variations expected under the hypothesis of linear
variation until the present day over a period as short as a 
year~\cite{Prestage:1995zz,Karshenboim:2000rg,Marion:2002iw}.

Another consequence of a variation in the parameters relevant to hadron
structure is the possibility of observable consequences in big bang
nucleosynthesis (BBN) or other nuclear phenomena such as the
composition of the ash of long extinct natural nuclear 
reactors~\cite{Shlyakhter_76,Damour:1996zw,FOth}. 
In this context, the effect of quark mass changes on the nucleon-nucleon 
force has been studied in effective field theory~\cite{Beane:2002vs,Epelbaum:2002gb}, 
most recently including constraints from lattice QCD~\cite{Beane:2002xf,Soto:2011tb}.
{}For the moment these lattice studies are at too high a quark mass to 
provide an accurate constraint~\cite{Soto:2011tb,Young:2002ib}.
In an alternative approach based on a more traditional model of the 
nucleon-nucleon force, the 
latest work of Flambaum and Wiringa~\cite{FW} 
on this topic involved the study of the variation 
of nuclear binding with quark mass using the Argonne potential and 
Schwinger-Dyson estimates of the variation of meson masses. 

In this work we employ a one boson-exchange (OBE) model of the 
nuclear force to
calculate the variation with changes in the quark mass of the binding
energies of selected finite nuclei as well as the energy 
of the $^1$S$_0$ anti-bound state  and 
the binding energies of the deuteron and triton. 
Apart from its intrinsic interest, this approach complements the work 
of Ref.~\cite{FW} and a comparison of the two provides one 
way to gauge the possible model dependence of the variations reported.
The method used here involves a detailed study of the variation of the mass
of each of the mesons usually employed in a one-boson-exchange (OBE) picture
of the nucleon-nucleon (NN) force. Care is taken to estimate this shift at the
relevant kinematic point, not just at the real, on-shell meson mass
or its pole position. These changes are then introduced into the
quark-meson coupling (QMC) model for some light nuclei and a typical OBE
model for the two nucleon systems and a Faddeev calculation of the triton.

In section II we examine, in turn, each of the mesons $\sigma_0, \,
\sigma_1, \, \omega, \, \rho, \, \pi$ and $\eta$. Section III presents
results for finite nuclei, while the two nucleon system and triton 
are discussed in section IV.
The final section is reserved for some concluding remarks.

\section{Meson masses}
In order to find the variations in the mass of the exchanged bosons 
in the nuclear
interaction ($\sigma$, $\rho$, and $\omega$) with respect to changes in the
mass of the pion $m_{\pi}$, we use three ideas. First 
we introduce a description
of the bare mass of the $\sigma$ ($m_{\sigma}^{\left(  0\right)  }$) 
in terms of
$m_{\pi}$ using the Nambu-Jona-Laisino (NJL) 
model~\cite{Nambu:1961tp,Bentz:2001vc}. 
Second, we introduce
the contribution from the self energies for $\sigma$, $\rho$, and $\omega$.
{}Finally, we include these self energies in two different 
ways: for the $\sigma$
the loop diagram is fixed in such a way that the total propagator contains a
pole on the second sheet of the complex energy plane at the position found by
Leutwyler {\it et al.}~\cite{leutw1}; for the $\rho$ and $\omega$ we
use the chiral fit to partially quenched data from lattice QCD 
developed by Armour {\it et al.}~\cite{armour}.
{}Finally, through
the Gell-Mann-Oakes-Renner (GMOR) relation we relate those changes to
variations in the quark masses.

\subsection{VARIATION IN $m_{\sigma}$ WITH $m_{q}$}\label{sigmameson}
In this work we choose to parametrize the intermediate range attraction
in the nucleon-nucleon ($NN$) force in terms the exchange of a $\sigma$
meson, following the traditional one-boson-exchange (OBE) approach.
Earlier work on the effect of changes in quark masses by Flambaum and
Wiringa~\cite{FW}, used explicit two-pion exchange for this purpose.
Almost certainly the reality is somewhere in between these extremes and
a comparison between our results and those of Ref.~\cite{FW} should
serve to pin down the uncertainties in this sector of the calculation.

The existence of the $\sigma$ meson has been somewhat controversial, largely
because its width is comparable with its mass. However, a careful dispersion
relation treatment using the Roy equation has served to accurately locate 
a pole which can be unambiguously identified 
with the $\sigma$ meson. Of course,
because of the large imaginary part of the energy of this pole, one cannot
easily relate the position of the pole to the position of a bump in the $\pi \pi$ cross
section. When it comes to the mass of the virtual $\sigma$ meson exchanged
in a OBE $NN$ potential, it is a third value that is of interest. Indeed,
the invariant mass of a meson exchanged in a typical $NN$ interaction is
very near zero and so we actually need the sigma mass for $p^2 \sim 0$.
This is most readily found within an effective Lagrangian approach.
\begin{figure}[ht]
\includegraphics{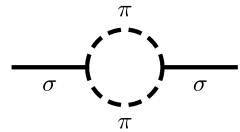}
\caption{Self-energy contributions for the $\sigma$ meson.}
\label{fig:1}
\end{figure}

Our model for the $\sigma$ meson involves a ``bare'' $\sigma$ meson
coupled to two pions. When required, the variation of the mass of this
bare state with quark mass will be calculated within the NJL model.
In this approach, the propagator of the dressed $\sigma$ is
described as a bare scalar propagator plus
an infinite series of contributions of the form shown in Fig.~\ref{fig:1}.
Calling this self-energy $\Sigma_{\pi\pi}^{\sigma}$, 
the total propagator can be
written as:
\begin{align*}
\Delta_{\sigma} &= \frac{i}{p^{2}-\left(  m_{\sigma}^{\left(  0\right)  }\right)
^{2}}\\
&+\frac{i}{p^{2}-\left(  m_{\sigma}^{\left(  0\right)  }\right)  ^{2}%
}\left(  i\Sigma_{\pi\pi}^{\sigma}  \right)  \frac
{i}{p^{2}-\left(  m_{\sigma}^{\left(  0\right)  }\right)  ^{2}}+\ldots,
\end{align*}
which resums to
\be
\label{propsigma}
\Delta_{\sigma}=\frac{i}{p^{2}-\left(  m_{\sigma}^{\left(  0\right)  }\right)
^{2}+\Sigma_{\pi\pi}^{\sigma}  }.
\ee
The pole in $\Delta_{\sigma}$ is the mass of the $\sigma$
resonance. This pole was calculated by Leutwyler {\it et al.} 
using the method of
Roy equations, which is model independent~\cite{leutw2}.
They obtained a pole located in the complex second sheet for $p$ at
\be
\label{poleleutw}
p=m_{\sigma}-\frac{i}{2}\Gamma 
= 441_{-8}^{+16}-i272_{-12.5}^{+9}~\text{MeV} \, .
\ee
The real part of the position of the resonance, $m_{\sigma}=441$ MeV, 
is in the
range $\left(400-1200\right)$ MeV 
given by the Particle Data Group (PDG) \cite{pdg},
while its width, $\Gamma=544$ MeV, is within 
the range $\left(  600-1000\right)$ MeV,
also from the PDG.

Having a reliable value for the $\sigma$ pole we can find a relation that lets
us fix $\Sigma_{\pi\pi}^{\sigma}$ such that
\be
\label{poleeq}
\sqrt{\left(  m_{\sigma}^{\left(  0\right)  }\right)  ^{2}-\Sigma_{\pi\pi
}^{\sigma}\left(  m_{\sigma}^{2}\right)  }\simeq441-i272~\text{MeV}.
\ee
With derivative coupling of the bare $\sigma$ to two pions
(consistent with chiral symmetry),
the expression for the $\pi \pi$ self-energy is found to be:
\begin{align}
\label{selfsigma}
i\Sigma_{\pi\pi}^{\sigma}=\frac{3}{2}\gamma^{2}_0\int
\frac{d^{4}k}{\left(  2\pi\right)  ^{4}}\frac{\left[  k^{\mu}\left(
p-k\right)  _{\mu}\right]  ^{2}}{\left(  k^{2}-m_{\pi}^{2}\right)  \left(
\left(  p-k\right)  ^{2}-m_{\pi}^{2}\right)  },
\end{align}
where $k$ represents the pion loop momentum, $p$ the $\sigma$
momentum and $\gamma$ the $\sigma \pi \pi$ coupling 
(initially we took the value
$\gamma_{0}$ from Harada, Sannino, and Schechter~\cite{lagrangian}).
We are considering all these particles as elementary,
so this is just an effective theory, and like any other effective theory
it has to be regularized. The regularization scheme we choose is
to impose a dipole cut-off (at each vertex) on the loop momentum
with mass $\Lambda$:
\be
\label{regulator}
\left[  1-\frac{\left(  \frac{p}{2}-k\right)  ^{2}}{\Lambda^{2}}\right]
^{-4} \, ,
\ee
which is sufficient to ensure convergence.
This dipole regulator
contains simple poles in $k$ 
(after writing it in the form of derivatives with
respect to $\Lambda$), which permits us to use contour integration
over the time component.

{}For the remaining integral over the three-momentum 
we rotated $\vec{k}$ in the complex plane $\left\vert
\vec{k}\right\vert e^{i\theta}$, with $-\frac{3\pi}{2}<\theta<0$, to ensure
that the imaginary part is located in the complex second Riemann sheet.
We also
performed a numerical integration with the help of the
routine \textit{NIntegral} of \textit{Mathematica}.
The final value of $\Sigma_{\pi\pi}^{\sigma}\left(
p^{2}\right) $ depends on two parameters:
the regularization mass $\Lambda$,
and the coupling constant $\gamma_0$.
We choose a range of values for $\Lambda$ such that, after fixing $\gamma_0$ and $m_{\sigma}^{(0)}$ to reproduce the pole position (Eq.~\eqref{poleleutw}), $m_{\sigma}^{(0)}$ varies from $560$ to $950$ MeV.
The results are summarized in Table~\ref{TabSigma1},
where $\Delta m_{\sigma}$ represents the deviation of our
result for the pole position from that of Leutwyler and collaborators.
\begin{table}
\caption{Parameters fixed to reproduce the position
of the $\sigma$ meson
pole ($\gamma_{0}=6.416\times10^{-3}$ $\left(\text{MeV}^{-1}\right)$). 
($\Delta m_\sigma$ is the deviation of the fitted from the empirical 
value.) }
\label{TabSigma1}
\begin{ruledtabular}
\begin{tabular}[c]{cccc}
\multicolumn{1}{c}{$\gamma$ $\left(\times\gamma_{0}\right)$} & \multicolumn{1}{c}{$\Lambda$ $\left(  MeV\right)  $}
& \multicolumn{1}{c}{$m_{\sigma}^{\left(0\right)  }$ $\left(  MeV\right)  $}
& \multicolumn{1}{c}{$\triangle m_{\sigma}~\left( MeV\right)\times10^{-5}  $}\\\hline
$4.56$ & $320.00$ & $563.54$ & $1.2-1.0i$\\
$4.60$ & $330.00$ & $600.23$ & $1.1-0.5i$\\
$4.70$ & $340.00$ & $639.85$ & $1.2-0.0i$\\
$4.83$ & $350.00$ & $683.46$ & $1.2-0.0i$\\
$5.02$ & $360.00$ & $732.00$ & $0.0+0.0i$\\
$5.27$ & $370.00$ & $790.18$ & $0.9-1.2i$\\
$5.61$ & $380.00$ & $859.15$ & $0.4-0.9i$\\
$6.07$ & $390.00$ & $945.43$ & $0.9-0.9i$\\
\end{tabular}
\end{ruledtabular}
\end{table}

We then define $m_{\sigma}^{2}\left(  OBE\right)
=\left(  m_{\sigma}^{\left(
0\right)  }\right)  ^{2}-\Sigma_{\pi\pi}^{\sigma}\left(  0\right)  $,
because
in an OBEP model for the $NN$ interaction the exchanged boson has nearly
zero momentum. $m_{\sigma}^{2}\left(  OBE\right)
$ is a real value because $\Sigma_{\pi\pi}^{\sigma}\left(
0\right)  $ is real. Thus any
variation on $m_{\sigma}\left(  OBE\right)  $ with respect to $m_{\pi}$
is given by variations in $m_{\sigma}^{\left(  0\right)  }$
and $\Sigma
_{\pi\pi}^{\sigma}\left(  0\right)  $:%
\be
\label{diffmobe}
\frac{\delta m_{\sigma}^{2}\left(  OBE\right)  }{\delta m_{\pi}^{2}}%
=\frac{m_{\sigma}^{\left(  0\right)  }}{m_{\pi}}\frac{\delta m_{\sigma
}^{\left(  0\right)  }}{\delta m_{\pi}}-\frac{\delta\Sigma_{\pi\pi}^{\sigma
}\left(  0\right)  }{\delta m_{\pi}^{2}}
\ee
and using the Gell-Mann-Oakes-Renner (GMOR) relation~\cite{gmor}:
\be
\label{dmsigmamq}
\frac{\delta m_{\sigma}\left(  OBE\right)  }{m_{\sigma}\left(  OBE\right)
}=\nu_{\sigma}\frac{\delta m_{q}}{m_{q}} \, ,
\ee
with
\be
\label{coeffmsigma}
\nu_{\sigma}=\frac{m_{\pi}^{2}}{2m_{\sigma}^{2}\left(  OBE\right)
}\left[  \frac{\delta\left(  m_{\sigma}^{\left(  0\right)  }\right)  ^{2}%
}{\delta m_{\pi}^{2}}-\frac{\delta\Sigma_{\pi\pi}^{\sigma}\left(  0\right)
}{\delta m_{\pi}^{2}}\right] \, .
\ee
We change $m_{\pi}$ near the physical value and
find the variation $\frac{\delta\Sigma_{\pi\pi}^{\sigma}\left(  0\right)
}{\delta m_{\pi}^{2}}$, which is almost constant for a small change in $m_{\pi
}^{2}$, so we only need to find the slope of the plot $\Sigma_{\pi\pi}^{\sigma
}\left(  0\right)  $ versus $m_{\pi}^{2}$ . We also need the variation of
$m_{\sigma}^{\left(  0\right)  }$ with $m_{\pi}$ near 
140 MeV. For this purpose we 
used the NJL model, which is known
to respect the chiral behaviour of QCD, including the GMOR relation.
The results for all the cases in Table~\ref{TabSigma1}
are contained in Table~\ref{TabSigma2}.%
\begin{table}\label{TabSigma2}
\caption{Calculations for the coefficient $\nu_\sigma$, which relates the
fractional change of the mass of the $\sigma$ meson relevant to the OBEP
model to the fractional change in the quark mass.}
\label{TabSigma2}
\begin{ruledtabular}
\begin{tabular}[c]{ccccc}
\multicolumn{1}{c}{$m_{\sigma}^{\left(0\right)  }$}
& \multicolumn{1}{c}{$\frac{\delta\Sigma_{\pi\pi}^{\sigma}(0)}{\delta m_{\pi}^2}$}
& \multicolumn{1}{c}{$\frac{\delta\left(m_{\sigma}^{(0)}\right)^2}{\delta m_{\pi}^2}$}
& \multicolumn{1}{c}{$\frac{\delta m_{\sigma}^2(OBE)}{\delta m_{\pi}^2}$}
& \multicolumn{1}{c}{$\nu_\sigma$}\\\hline
$563.54$ & $-0.145$ & $2.677$ & $2.822$ & $0.089$\\
$600.23$ & $-0.164$ & $2.632$ & $2.796$ & $0.078$\\
$639.85$ & $-0.189$ & $2.576$ & $2.765$ & $0.068$\\
$683.46$ & $-0.220$ & $2.546$ & $2.766$ & $0.060$\\
$732.00$ & $-0.261$ & $2.502$ & $2.763$ & $0.052$\\
$790.18$ & $-0.314$ & $2.451$ & $2.765$ & $0.045$\\
$859.15$ & $-0.389$ & $2.401$ & $2.790$ & $0.038$\\
$945.43$ & $-0.495$ & $2.344$ & $2.839$ & $0.032$\\
\end{tabular}
\end{ruledtabular}
\end{table}

{}From Table~\ref{TabSigma2}, we notice that as $\Lambda$ increases
($\Lambda$ increases when $m_{\sigma
}^{\left(  0\right)  }$ grows) the value of $\Sigma_{\pi\pi
}^{\sigma}\left(  0\right)  $ also grows and its contribution of $\frac{\delta
\Sigma\left(  0\right)  }{\delta m_{\pi}^{2}}$ to $\frac{\delta m_{\sigma}%
^{2}\left(  OBE\right)  }{\delta m_{\pi}^{2}}$ increases. However,
in practice the variation of the $\sigma$ bare mass is much larger
and, in total, the larger
$m_{\sigma}^{\left(  0\right)  }$ the smaller the coefficient $\nu_\sigma$.
What is more, calculations within the Quark Meson Coupling (QMC) model
tend to favour values for $m_{\sigma}\left(  OBE\right)  $
near $550$ MeV (see sect.~\ref{QMC}).

\subsection{VARIATIONS IN $m_{\rho}$ AND $m_{\omega}$
WITH RESPECT TO $m_{q}$}
\label{rhoandomega}
In the case of the $\rho$ meson we have a good deal of data taken from
lattice calculations in
partially quenched QCD from the CP-PACS collaboration.
Armour {\it et al.}~\cite{armour} used
this data in an analysis that included the leading and next-to-leading
non-analytic chiral corrections to the self-energy to make
an extrapolation of
the mass $m_{\rho}$ to the chiral limit ($m_{\pi}\approx0$).
At the physical value of $m_{\pi}$ they found excellent
agreement with the physical value.

\begin{figure}[ht]
\includegraphics{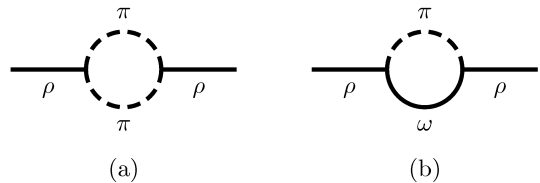}
\caption{Self-energy contributions for the $\rho$-meson.}
\label{fig:rho}
\end{figure}

The relevant self-energy diagrams for the $\rho$ are 
given in Fig~\ref{fig:rho}.\\
\\
These yield the following expressions:
\be
\label{selfrho}
\Sigma_{\pi\pi}^{\rho}=-\frac{f_{\rho\pi\pi}^{2}}{6\pi^{2}}\int_{0}^{\infty
}\frac{k^{4}u_{\pi\pi}^{2}\left(  k\right)  dk}{\omega_{\pi}\left(  k\right)
\left(  \omega_{\pi}^{2}\left(  k\right)  -\frac{\mu_{\rho}^{2}}{4}\right)  }%
\, , 
\ee
\be
\label{selfrhoomega}
\Sigma_{\pi\omega}^{\rho}=-\frac{g_{\omega\rho\pi}}{12\pi^{2}}\mu_{\rho}%
\int_{0}^{\infty}\frac{k^{4}u_{\pi\omega}^{2}\left(  k\right)  dk}{\omega
_{\pi}^{2}\left(  k\right)  }%
\, , 
\ee
where $f_{\rho\pi\pi}=6.028$ and $g_{\omega\rho\pi}=0.016~\text{MeV}^{-1}$. The
regularization functions used in the analysis are:
\be
\label{regulator1}
u_{\pi\omega}\left(  k\right)  =\frac{\Lambda^{4}}{\left(  \Lambda^{2}%
+k^{2}\right)  ^{2}} \, , 
\ee
\be
\label{regulator2}
u_{\pi\pi}\left(  k\right)  =\frac{\left(  \Lambda^{2}+\frac{\mu_{\rho}^{2}%
}{4}-\mu_{\pi}^{2}\right)  ^{2}}{\left(  \Lambda^{2}+k^{2}\right)  ^{2}}
\, , 
\ee
and we use the approximations: $m_{\pi}\ll m_{\omega,\rho}$ and
$m_{\rho}\approx m_{\omega}$.

The fit to the partially quenched lattice QCD data for the $\rho$ meson
involved a fit of the form:
\be
\label{mrhoexpan}
m_{\rho}=\sqrt{\left(a_{0}+a_{2}m_{\pi}^{2}+a_{4}m_{\pi}^{4}\right)^{2}+\Sigma_{TOT}}
\, ,
\ee
where $\Sigma_{TOT}=\Sigma_{\pi\pi}^{\rho}+\Sigma_{\pi\omega}^{\rho} $,
and the
coefficients, $a_i$, are: $a_{0}=832.00~\text{MeV}$, $a_{2}%
=4.94\times10^{-4}~\text{MeV}^{-1}$, $a_{4}=-6.10\times10^{-11}~\text{MeV}^{-3}$ and $\Lambda=655.00~\text{MeV}$ (up to
errors). At the physical pion mass (in full QCD) this
yields a value of:
\be
\label{mrhoexper}
m_{\rho}\approx778~\text{MeV} \, ,
\ee
which shows remarkable agreement with the physical value,
with a shift of only:
\be
\label{deltamrho}
m_{\rho}-m_{\rho}^{\rm phys}\sim3.7~\text{MeV} \, .
\ee

As in the case of the $\sigma$ meson, we consider a one boson exchange
potential
with almost zero momentum transfer, so that $\mu_{\rho}\sim0$
in the propagator of \eqref{selfrho} (not in the regulator,
because the mass that
appears there is the physical mass):
\be
\label{selfrhosimp}
\Sigma_{\pi\pi}^{\rho}=-\frac{f_{\rho\pi\pi}^{2}}{6\pi^{2}}\int_{0}^{\infty
}\frac{k^{4}u_{\pi\pi}^{2}\left(  k\right)  dk}{\omega_{\pi}^{3}\left(  k\right)
} \, .
\ee
This of course changes the value of $m_{\rho}$, now denoted 
$m_\rho({\rm OBE})$, at the physical
pion mass. Indeed, in
this case it is near $762$ MeV. The relation between $m_{\rho}$ and
$m_{\pi}^{2}$ near the physical value is almost linear and it presents a slope
of $\frac{\delta m_{\rho}({\rm OBE})}{\delta m_{\pi}^{2}}
= 0.00135~\text{MeV}^{-1}$. Following
the analysis for $m_{\sigma}$ we can relate this change with $m_{q}$ in the
following way:
\be
\label{dmrhodmq}
\frac{\delta m_{\rho}({\rm OBE})}{m_{\rho}}=
\left( \frac{m_{\pi}^{2}}{m_{\rho}({\rm OBE})}
\frac{\delta m_{\rho}({\rm OBE})}{\delta m_{\pi}^{2}}\right)  
\frac{\delta m_{q}%
}{m_{q}} \, .
\ee
Using for $m_{\rho}({\rm OBE)}$, henceforth simply written as 
$m_\rho$, the  value 770 MeV, 
which is usually used in OBE models, we find:
\be
\label{coeffmrho}
\frac{\delta m_{\rho}}{m_{\rho}}=0.034\frac{\delta m_{q}}{m_{q}}%
\, .
\ee

The analysis for the $\omega$ meson is closely related to that
of the $\rho$ meson. However, the diagrams
that contribute to the self-energy terms differ because there is no
two-pion contribution because of G-parity. In
addition, $\Sigma_{\pi\rho}^{\omega}$ is $3\times\Sigma_{\omega\pi}^{\rho}$,
because there are three possible $\rho - \pi$ charge combinations.
{}For the analytic terms in the expansion we use the
same coefficients ($a_i$ ) as in the case of the $\rho$,
because the mass difference between them is only of order $10$\ MeV.
Then the variation of $m_{\omega}$ with respect to $m_{\pi}^{2}$ near the
physical value gives $\frac{\delta m_{\omega}}{\delta m_{\pi}^{2}%
}=0.00096$ MeV$^{-1}$, which leads to the relation:
\be
\label{coeffmomega}
\frac{\delta m_{\omega}}{m_{\omega}}=0.024\frac{\delta m_{q}}{m_{q}} \, , 
\ee
where we used the physical mass for the $\omega$, $m_{\omega}=782$ MeV 
-- again because that is the value typically used in a OBE potential. 
(The value obtained at zero momentum transfer would be $765$ MeV.)

\subsection{SUMMARY OF MESON MASS VARIATION }
\begin{table}[htb]
\begin{center}
\caption{Coefficients $\nu_i$ summarising the rate of variation of 
the masses of the mesons used in an OBE description of the NN force 
with respect to quark mass - see Eq.~(\ref{vary}).}
\label{meson_summary}
\vspace{0.3 cm}
\begin{tabular}{lc}\hline
 Meson &~~~ $\nu$ (MeV) \\ \hline \hline
 $\pi$             & 0.5  \\ \hline
 $\eta$           &  0.012 \\ \hline
 ${\sigma_0}$ &    0.089 \\ \hline
 ${\sigma_1}$ &  0.072  \\ \hline
 $\rho$          & 0.034  \\ \hline
 $\omega$     &  0.024  \\ \hline
\end{tabular}
\end{center}
\end{table}
For the $\eta$, like the pion, we use the GMOR relation to 
calculate the variation within respect to $u$ and $d$ mass.
In the case of the iso-vector scalar meson, $\sigma_1$, 
which has negative G-parity and therefore does not couple to two pions, 
we use the NJL model - corresponding to the third column 
and second row of Table~\ref{TabSigma2}, 
and Eq.~(\ref{coeffmsigma}) without the self-energy part. 
{}For convenience, in Table~\ref{meson_summary} we summarise 
the values of $\nu_i$, 
defined as
\begin{equation}
\frac{\delta m_i}{m_i} = \nu_i \frac{\delta m_q}{m_q} \, , 
\label{vary}
\end{equation}
which will be used below.

\section{Nucleon Mass}
In order to compute the variation of nuclear binding energies with quark mass, we also need to know how the nucleon mass changes.

The variation with light quark mass is directly given by the so-called $\pi N$ sigma commutator
\be
\sigma_{\pi N}=m_{q}\langle N \mid \overline{q}q \mid N \rangle = m_{q}\frac{\delta m_{N}}{\delta m_{q}},
\ee
where $\overline{q}q\equiv \overline{u}u+\overline{d}d$.

The last equality, which gives the information we need, follows from the Feynman-Hellmann theorem. A number of methods have been used to extract $\sigma_{\pi N}$ from pion-nucleon scattering data using dispersion relations, but the resulting value is still controversial.

Instead, the most reliable method seems to be to use fits to lattice QCD data for $m_{N}$ as a function of $m_{q}$~\cite{Young}. These fits, which build the constraints of chiral effective field theory, appear to yield very reliable values. We take the result of the latest analysis of PACS-CS data by Shanahan et. al.~\cite{Shanahan}, namely $\sigma_{\pi N}=45 \pm 6$ MeV. Thus we use:
\be
\frac{\delta m_{N}}{m_{N}}=0.048\frac{\delta m_{q}}{m_{q}}.
\label{varmN}
\ee
\section{$^7$Li, $^{12}$C and $^{16}$O nuclei}\label{QMC}
To study the effect of the quark mass variation on the single-particle
energies of $^7$Li, $^{12}$C and $^{16}$O nuclei,
it is highly desirable to use a nuclear model based on the quark 
degrees of freedom.
The quark-meson coupling (QMC) model, 
which originated with Guichon~\cite{Guichon} as a description of 
nuclear matter and was extended and improved 
to describe the properties of finite
nuclei~\cite{QMCFinite1,QMCFinite2},
is ideal for this purpose.
The successful features of the QMC model applied to various nuclear
phenomena and hadronic properties in a nuclear medium, are reviewed
extensively in Ref.~\cite{QMCReview}.
The model has been updated to study the properties of
hypernuclei~\cite{QMCHyp},
and neutron star structure~\cite{QMCNeutronStar,Whittenbury:2012rn},
where the quark structure of the nucleons and hyperons should play
an important role at such high density.
We calculate the change in the single-particle energies of these nuclei
versus the current quark mass ($m_q$) and the mass 
of the nucleon ($m_N$) 
using the theory presented in Ref.~\cite{QMCFinite2} and the 
meson and nucleon mass changes calculated above.

In Ref.~\cite{QMCFinite2} the standard values used to reproduce the
nuclear matter saturation properties are,
($m_N,m_\sigma,m_\omega,m_\rho$)=(939,550,783,770) MeV, with the current
quark mass $m_q=5.0$ MeV. For the calculation of finite nuclei, 
the ratio for the $\sigma-N$ coupling constant and the mass,
($g^N_\sigma/m_\sigma$), was kept constant and fitted to the rms 
charge radius of $^{40}$Ca,
$r_{ch}$($^{40}$Ca) = 3.48 fm, by adjusting 
$m_\sigma \to \tilde{m}_\sigma$ = 418 MeV. (Note that 
the variation of $m_\sigma$ at fixed ($g^N_\sigma/m_\sigma$) has no effect
on the nuclear matter properties.) 
To account for this, 
we calculate the shift in $m_\sigma$ from 550 MeV for a given 
variation of the quark mass. This changes the ratio   
($g^N_\sigma/m_\sigma$) and from that new value we deduce the 
corresponding shift in $m_\sigma$ from 418 MeV, to be used in the finite
nucleus calculation. 
First, with the nucleon mass fixed at $m_N=939$ MeV and the 
variations of the meson masses,
$\delta m_{\sigma,\omega,\rho}$, evaluated for quark mass variations of 
$\delta m_q=\pm 0.05$ and $\pm 0.1$ MeV,
we calculate the single-energies in $^7$Li, $^{12}$C and $^{16}$O.
Note that the very small differences for the $\sigma$ and $\omega$ meson 
mass values
used to extract the relation in terms of $\delta m_q$ 
in~\ref{sigmameson} and~\ref{rhoandomega},
were neglected.
In addition, we also calculate the energy per nucleon (E/nucleon).
The results are given in Table~\ref{TabEnergy}.

\begin{table*}
\begin{center}
\caption{
Single-particle energies (in MeV) for $^7$Li, $^{12}$C and $^{16}$O 
nuclei versus
quark mass $m_q$ (in MeV) calculated in 
the quark-meson coupling (QMC) model~\cite{QMCFinite2}.
E/nucleon stands for energy per nucleon. The standard value for 
the quark mass
use in the QMC model is $m_q=5.00$ MeV.
}
\label{TabEnergy}
\begin{tabular}[t]{lllccccc}
\hline\hline
   &States &$m_q$
   &\hspace{2em}4.90\hspace{2em} &\hspace{2em}4.95\hspace{2em} &\hspace{2em}5.00\hspace{2em}
   &\hspace{2em}5.05\hspace{2em} &\hspace{2em}5.10\hspace{2em} \\
\hline\hline
$^7$Li & & & & & & & \\
\hline
 p &$1s_{1/2}$ &  &-19.0216 &-18.8657 &-18.7089 &-18.5522 &-18.3953 \\
   &$1p_{3/2}$ &  &-3.5945  &-3.5104  &-3.4267  &-3.3432  &-3.2602  \\
 n &$1s_{1/2}$ &  &-18.3503 &-18.2146 &-18.0777 &-17.9408 &-17.8039 \\
   &$1p_{3/2}$ &  &-3.2385  &-3.1709  &-3.1036  &-3.0366  &-2.9699  \\
\hline
   &E/nucleon & &-1.710 &-1.662 &-1.614 &-1.566 &-1.519\\
\hline\hline
$^{12}$C & & & & & & & \\
\hline
 p &$1s_{1/2}$ &  &-26.2579 &-26.1110 &-25.9643 &-25.8179 &-25.6716 \\
   &$1p_{3/2}$ &  &-10.0448 &-9.9445  &-9.8444  &-9.7447  &-9.6453  \\
 n &$1s_{1/2}$ &  &-29.4077 &-29.2558 &-29.1040 &-28.9525 &-28.8011 \\
   &$1p_{3/2}$ &  &-12.9435 &-12.8379 &-12.7327 &-12.6277 &-12.5231 \\
\hline
   &E/nucleon & &-4.174 &-4.107 &-4.040 &-3.974 &-3.908\\
\hline\hline
$^{16}$O & & & & & & & \\
\hline
 p &$1s_{1/2}$ &  &-29.0713 &-28.9260 &-28.7810 &-28.6362 &-28.4917 \\
   &$1p_{3/2}$ &  &-13.8605 &-13.7503 &-13.6405 &-13.5309 &-13.4217 \\
   &$1p_{1/2}$ &  &-12.0635 &-11.9647 &-11.8661 &-11.7679 &-11.6700 \\
 n &$1s_{1/2}$ &  &-33.0861 &-32.9358 &-32.7857 &-32.6358 &-32.4862 \\
   &$1p_{3/2}$ &  &-17.6266 &-17.5112 &-17.3962 &-17.2815 &-17.1671 \\
   &$1p_{1/2}$ &  &-15.8151 &-15.7110 &-15.6073 &-15.5038 &-15.4006 \\
\hline
   &E/nucleon & &-6.109 &-6.035 &-5.961 &-5.888 &-5.815\\
\hline\hline
\end{tabular}
\end{center}
\end{table*}
%
\begin{table*}
\begin{center}
\caption{
Single-particle energies (in MeV) for $^7$Li, $^{12}$C and $^{16}$O
nuclei versus
nucleon mass $m_N$ (in MeV), calculated in the
quark-meson coupling (QMC) model~\cite{QMCFinite2}.
E/nucleon stands for energy per nucleon. The standard value for
the nucleon mass
used in the QMC model is $m_N=939.0$ MeV.
}
\label{TabEnergyN}
\begin{tabular}[t]{lllccccc}
\hline\hline
   &States &$m_N$
   &\hspace{2em}938.0\hspace{2em} &\hspace{2em}938.5\hspace{2em} &\hspace{2em}939.0\hspace{2em}
   &\hspace{2em}939.5\hspace{2em} &\hspace{2em}940.0\hspace{2em} \\
\hline\hline
$^7$Li & & & & & & & \\
\hline
 p &$1s_{1/2}$ &  &-18.6768 &-18.6929 &-18.7089 &-18.7249 &-18.7409 \\
   &$1p_{3/2}$ &  &-3.4010  &-3.4139  &-3.4267  &-3.4396  &-3.4524  \\
 n &$1s_{1/2}$ &  &-18.0517 &-18.0647 &-18.0777 &-18.0907 &-18.1036 \\
   &$1p_{3/2}$ &  &-3.0830  &-3.0933  &-3.1036  &-3.1138  &-3.1241  \\
\hline
   &E/nucleon & &-1.600 &-1.607 &-1.614 &-1.621 &-1.628\\
\hline\hline
$^{12}$C & & & & & & & \\
\hline
 p &$1s_{1/2}$ &  &-25.9445 &-25.9544 &-25.9643 &-25.9742 &-25.9841 \\
   &$1p_{3/2}$ &  &-9.8211  &-9.8328  &-9.8444  &-9.8561  &-9.8677  \\
 n &$1s_{1/2}$ &  &-29.0828 &-29.0934 &-29.1040 &-29.1147 &-29.1253 \\
   &$1p_{3/2}$ &  &-12.7076 &-12.7202 &-12.7327 &-12.7452 &-12.7577 \\
\hline
   &E/nucleon & &-4.022 &-4.031 &-4.040 &-4.050 &-4.059\\
\hline\hline
$^{16}$O & & & & & & & \\
\hline
 p &$1s_{1/2}$ &  &-28.7634 &-28.7722 &-28.7810 &-28.7897 &-28.7985 \\
   &$1p_{3/2}$ &  &-13.6182 &-13.6293 &-13.6405 &-13.6516 &-13.6627 \\
   &$1p_{1/2}$ &  &-11.8432 &-11.8547 &-11.8661 &-11.8775 &-11.8889 \\
 n &$1s_{1/2}$ &  &-32.7665 &-32.7761 &-32.7857 &-32.7952 &-32.8048 \\
   &$1p_{3/2}$ &  &-17.3721 &-17.3841 &-17.3962 &-17.4082 &-17.4202 \\
   &$1p_{1/2}$ &  &-15.5826 &-15.5949 &-15.6073 &-15.6196 &-15.6319 \\
\hline
   &E/nucleon & &-5.941 &-5.951 &-5.961 &-5.971 &-5.981\\
\hline\hline
\end{tabular}
\end{center}
\end{table*}
%

{}From Table~\ref{TabEnergy} we see that the absolute values 
of the single-particle binding 
energies of each nucleus
decrease as the quark mass increases.
This is because an increase of the quark mass 
leads to a significant increase of the mass of the $\sigma$ meson
and this reduces the attraction arising from $\sigma$ meson exchange 
by more than the repulsion associated with the $\omega$ decreases. 
It is interesting to point out that a small variation of the
quark mass of 0.05 MeV is reflected in a change in 
the single-particle energies
of order of 0.1 MeV. That is, the impact is appreciable.
Furthermore, we note that the binding energy per nucleon for
each nucleus decreases linearly as the quark
mass increases.

Next, we calculate the variation of the 
single-particle energies as the mass of the nucleon is varied.
The results are 
given in Table~\ref{TabEnergyN} for the same nuclei as 
in Table~~\ref{TabEnergy}.
As the value of the nucleon mass increases 
the absolute values of the single-particle
binding energies also increase. 
This seems to be natural, since the kinetic energy
is suppressed.

It may be helpful to consider the 
binding energy per nucleon as a function of the quark mass.
Based on the results given in Tables~\ref{TabEnergy} 
and~\ref{TabEnergyN}, and in \eqref{varmN};
we get the following relations for each nucleus:
%
\bg
\frac{\delta \left|{\rm E}_{^7{\rm Li}}\right|/{\rm nucleon}}{\left|{\rm E}_{^7{\rm Li}}\right|/{\rm nucleon}} 
&=& -2.571\frac{\delta m_{q}}{m_{q}},\\  
\label{ELi7}\notag\\
\frac{\delta \left|{\rm E}_{^{12}{\rm C}}\right|/{\rm nucleon}}{\left|{\rm E}_{^{12}{\rm C}}\right|/{\rm nucleon}} 
&=& -1.438\frac{\delta m_{q}}{m_{q}},\\
\label{EC12}\notag\\
\frac{\delta \left|{\rm E}_{^{16}{\rm O}}\right|/{\rm nucleon}}{\left|{\rm E}_{^{16}{\rm O}}\right|/{\rm nucleon}} 
&=& -1.082\frac{\delta m_{q}}{m_{q}}.
\label{EO16}
\en
The contributions to the previous coefficients from the variation of the exchanged mesons masses are found from Table~\ref{TabEnergy}, and from Table~\ref{TabEnergyN} we obtain the contribution from the nucleon mass. These calculations are summarised in the following equation:\\
\be
\frac{\delta \left|{\rm E}_{i}\right|/{\rm nucleon}}{\left|{\rm E}_{i}\right|/{\rm nucleon}}=\left(\nu_{mesons}+\nu_{Nucleon}\right)\frac{\delta m_{q}}{m_{q}}\notag, 
\ee
with $i$ representing each of the three nuclei we are considering, $\nu_{mesons}$ being described by
\be
\nu_{mesons}=\frac{\delta \left|{\rm E}_{i}\right|/{\rm nucleon}}{\delta m_{q}}\cdot\frac{m_{q}}{\left|{\rm E}_{i}\right|/{\rm nucleon}}\notag, 
\ee
and $\nu_{Nucleon}$ by
\be
\nu_{Nucleon}=\frac{\delta \left|{\rm E}_{i}\right|/{\rm nucleon}}{\delta m_{N}}\cdot\frac{m_{N}}{\left|{\rm E}_{i}\right|/{\rm nucleon}}\cdot0.048\notag.
\ee
%

\section{Variation in the energies of the two- and three-nucleon system 
with variation in the meson and nucleon masses}\label{deut}
To examine the variation in the binding energy of the deuteron and 
triton with changes in the meson and nucleon masses, we need to consider 
a purely One Boson Exchange (OBE) model for the nucleon-nucleon 
interaction. We choose to employ the OBE potential of 
Bryan-Scott (BS)~\cite{BS69}, which includes the exchange 
of $(\pi, \eta, \sigma_0, \sigma_1, \rho, \omega)$-mesons. 
To avoid the singular nature of this potential, BS introduced a monopole 
regularization scheme that insured that the potential is finite 
at the origin. With a cutoff mass of 1500~MeV this regularization 
is shorter in range than the range of the heaviest of the bosons 
included in the potential. As a result, the medium range 
interaction is dominated by the $\sigma_0$, $\sigma_1$ followed 
by the $\rho$ and $\omega$ exchanges. 

Because of the nonlocal nature of the potential (term proportional 
to $\nabla^2$), we have used the method of moments~\cite{AR75} to 
solve the Schr\"odinger equation for the binding energy of 
the deuteron and the $^1$S$_0$ amplitude. 
This entails expanding the radial wave function 
$\psi_\ell(r)$ for a given angular momentum $\ell$ as a linear combination 
of Yamaguchi~\cite{YY54} wave functions $\psi_\ell^{(Y)}(r;\beta_i)$ 
with different range parameters $\beta_i$, \textit{i.e.}~\cite{AG12}
\be
\psi_\ell(r) = \sum_{i=1}^n\ b_i^\ell\ \psi^{(Y)}_\ell(r;\beta_i)\ ,
\label{eq:5.1}
\ee
where we have taken $n=12$ and the $\beta_i$ are multiples of the pion mass. 
The  present choice for the variational wave function ensures that the 
correct long-range behavior of $\psi_\ell(r)$ is that defined 
by the asymptotic behavior of $\psi^{(Y)}_\ell(r)$. This in turn is 
determined by the binding energy of the deuteron or 
the $^1$S$_0$ anti-bound state. This procedure 
reduces the two-body Schr\"odinger equation to a set of $2n$ 
homogenous algebraic equations that give us the binding energy and 
the wave function for the deuteron to a very good 
approximation~\cite{AR75,AG12}.

For the $^1$S$_0$, the pole in the scattering amplitude is on the 
second energy sheet, and the analytic continuation of the method of 
moments to the second energy sheet is not as simple, 
because the pole is 
along the negative imaginary momentum axis. However, since this 
anti-bound state pole is close to the zero energy ($E_{P}=-0.066$~MeV), 
we have chosen the zero energy point to reduce 
the Schr\"odinger equation using the method of moments to a set 
of $n$ algebraic equations. It has been demonstrated~\cite{AG12} 
that this procedure gives a good representation of the original 
potential for the low energy scattering parameters. 
As a result we use the effective range expansion to determine 
the position of the anti-bound state pole in the momentum or 
$k$-plane, \textit{i.e.} we write the on-shell $^1$S$_0$ amplitude 
in terms of the phase shifts $\delta_0$ as
\begin{equation}
t(k) = -\frac{\hbar^2}{\pi\mu}\ \frac{1}{k\cot\delta_0 - i k} \ ,\label{eq:5.2}
\end{equation}
where $\mu$ is the reduced mass, 
and make use of  the effective range expansion 
\begin{equation}
k\cot\delta_0 = -\frac{1}{a_s} + \frac{1}{2}\,r_s\,k^2 - P_s\,r_s^3\,k^4 + \cdots\ ,\label{eq:5.3}
\end{equation}
where $P_s$ is the shape parameter, to analytically continue 
the amplitude onto the second energy sheet. 
Since the anti-bound state is close to zero energy ($k\approx - 0.04\,i$), 
we can truncate the effective range expansion to include the $k^4$ term. 
To test the accuracy of this procedure, we compare the position of 
the pole on the second energy sheet for the Bryan-Scott potential 
by truncating at $k^2$ and $k^4$ term with the result 
$E_{P}= -0.0711531$ and $-0.0711548$~MeV respectively. 
As a result we have chosen to truncate the effective range 
expansion to include the $k^4$ term.

The use of the trial function in Eq.~(\ref{eq:5.1}) has the added 
advantage of allowing us to construct an equivalent rank one 
separable potential, often referred to as the Unitary Pole 
Approximation (UPA),  that has identically the same deuteron wave function 
as the original OBE potential~\cite{AR75}. After partial wave expansion, 
this is of the form
\be
V^{\rm UPA}_{\ell;\ell'}(k,k') = g_\ell(k)\,C_{\ell;\ell'}\,g_{\ell'}(k')\ ,\label{eq:5.4}
\ee
where the form factors are directly related to the radial wave function 
$\psi_\ell(r)$ and the strength of the potential is adjusted to insure 
that the matrix element of the UPA and original OBE potential are identical 
at the energy of the pole in the amplitude. 
The same procedure is applied to the $^1$S$_0$ channel.

\begin{table}[htb]
\begin{center}
\caption{Variation in the position of the anti-bound state pole on the second energy sheet, the binding energy of the deuteron and triton
with changes in hadron mass $m_H$. For the Bryan-Scott potential the position of the anti-bound state pole is $E_P$ = -7.1155E-02~MeV, 
deuteron binding energy $E_D= 2.18365$~MeV, while the triton binding
energy $E_t= 7.9131$~MeV in the UPA.}
\label{tablebind}
\vspace{0.3 cm}
\begin{tabular}{lcccc}\hline
 H &~~~ $m_H$ (MeV) ~~& $\frac{\delta E_P}{\delta m_H}$ &~~~ $\frac{\delta E_D}{\delta m_H}$ ~~~&~~~~ $\frac{\delta E_t}{\delta m_H}$ ~~~ \\ \hline \hline
 $\pi$             & 138.7   & 2.38E-03 &  - 0.0201   &  -0.0146  \\ \hline
 $\eta$           & 548.7  &  -7.40E-05 &     0.0019   &   0.0034  \\ \hline
 ${\sigma_0}$ & 550.0  & -1.99E-03&    -0.1026   &  -0.3355   \\ \hline
 ${\sigma_1}$ & 600.0  &  -3.09E-03&    0.0486   &   0.0790   \\ \hline
 $\rho$          & 763.0   &  1.27E-04&   -0.0295   &  -0.0517   \\ \hline
 $\omega$     & 782.8   &  1.34E-02&  0.0923    &  0.2776    \\ \hline
 $N$             & 938.92  & 2.95E-04  & 0.0289   &  0.0527    \\ \hline\hline
\end{tabular}
\end{center}
\end{table}

Having constructed a rank one separable potential equivalent to the 
OBE potential, we can write the Faddeev equations as a set of coupled 
one dimensional integral equations~\cite{AT77}. If one includes the 
$^1$S$_0$ and $^3$S$_1$-$^3$D$_1$ nucleon-nucleon partial waves only, 
then the number of coupled integral equations reduces to five, 
and these can be solved for the binding energy and wave function 
of the triton~\cite{AB77}. 

To examine the variation in the binding energy with changes in the mass 
of the mesons and nucleon, we have calculated the slope of the binding 
energy as function of the mass at the value of the mass used in the 
OBE potential. 
In Table~\ref{tablebind} we present this variation in the energy of anti-bound state, the deuteron and 
triton binding energies with respect to the variation in the masses 
of the six bosons included in the OBE potential. 
We have also included the variation in the binding energies with changes 
in the nucleon mass $m_N$.
Here, we note that the nucleon mass is present, 
not only in the kinetic energy of the two- and three-body equations, but also in the 
definition of the Bryan-Scott OBE potential. For the one pion exchange component, the strength of the potential is proportional to $(g_{\pi NN}/2M)^2$ which is equivalent to $(f_{\pi NN}/m_\pi)^2$ had BS used a pseudo-vector coupling in the Lagrangian. From the Goldberger-Treiman~\cite{GT58} relation we have that
\begin{equation}
\frac{g_{\pi NN}}{M} \propto \frac{g_A}{f_\pi}\ ,\label{eq:5.5}
\end{equation}
where $f_\pi$ is the pion decay constant. Although $g_A$ and $f_\pi$ are dependent on the quark mass, the ratio to first order is not sensitive to variation in quark mass. This suggests that the strength of the one pion exchange component of the BS should not change with changes in the nucleon mass. Since the $\eta$ is part of the same $SU(3)$ octet as the pion, one could apply the same argument the $\eta$ exchange component of the OBE potential. For the scalar ($\sigma_0$ and $\sigma_1$)  and vector ($\rho$ and $\omega$) meson exchanges, the relative strength of the central, the spin-orbit and the tensor component depend on the nucleon mass, and to that extent, we have maintained the $M$ dependence of the OBE potential for the scalar and vector exchanges.
 From Table~\ref{tablebind} 
we observe that the variation is largest for the $\sigma_0$ and $\omega$, 
{}followed by the variation with the $\pi$, $\sigma_1$, $\rho$ and $N$ masses, 
with the variation in the energy with the $\eta$ mass being minimal.

\subsection{Total variation in binding energies}
From the detailed results given in Table~\ref{tablebind} and the 
earlier results for the variation of the meson and nucleon masses 
with quark mass, we can readily 
deduce the total variation of the deuteron and triton binding energies 
and the energy of the anti-bound state, $E_{P}$,  
with changes in the quark mass:  
\be
\frac{\delta E_D}{E_D} = -0.912 \frac{\delta m_q}{m_q} \, ,
\label{eq:D_total}
\ee
\be
\frac{\delta E_t}{E_t} = -0.980 \frac{\delta m_q}{m_q} 
\label{eq:T_total}
\ee
and
\be
\frac{\delta E_P}{E_P} = -2.839 \frac{\delta m_q}{m_q} \, .
\label{eq:Ep_total}
\ee
The details for these calculations are shown in the Appendix.

The variations of the deuteron and triton binding energies given in 
Eqs.~(\ref{eq:D_total}) and (\ref{eq:T_total}), respectively, are 
completely compatible with those reported by Flambaum and 
Wiringa~\cite{FW}. In particular, the coefficiants on the rhs of those 
equations, namely -0.91 for the deuteron and -0.98 for the triton, 
are very close to those reported in Ref.~\cite{FW} for the AV14 potential, 
namely -0.84 and -0.89.

On the other hand, for the $^1S_0$ anti-bound state, with energy $E_P$, there 
is a significant disagreement. The sign reported above for $\delta E_P/E_P$ 
is negative, whereas a positive value was reported in Ref.~\cite{FW}. 
Since Dmitriev {\it et al.}~\cite{FW} presented an apparently general 
argument relating the change in the deuteron binding to that in the 
energy of the anti-bound state, we re-checked every term in our calculation 
carefully. There is no doubt that our result is correct for the model 
used. We note that, from Table II of Flambaum and Wiringa~\cite{FW}, 
the individual pieces of the Argonne potential do {\it not} respect 
the supposedly general result of Dmitriev {\it et al.} and therefore 
it cannot be a model independent result. We note, in particular, that 
the tensor force plays a significant role for the deuteron, whereas it 
is absent for the $^1S_0$ channel. Clearly, this difference for the 
$^1S_0$ anti-bound state will lead to significant changes when one computes 
the effect of  a change in quark mass on the reaction rate for $n \, p 
\rightarrow d \, \gamma$.

\section{CONCLUSIONS}
\label{conclusion}
We have calculated the variation of the binding energy of the deuteron,  
triton  and the $^1$S$_0$ anti-bound pole position, 
as well as the binding energy per nucleon for a number of 
light nuclei, with respect to variations in the light (average of $u$ and $d$) 
quark mass. The results,  
expressed in terms of a parameter $K_A$, defined by
\begin{equation}
\frac{\delta BE(A)}{BE(A)} = K_A \frac{\delta m_q}{m_q} \, ,
\label{eq:KA}
\end{equation}
are summarised in Table-\ref{coeff_summary}. In order to 
determine these coefficients, we first calculated the change with 
quark mass of the mesons used in a typical one-boson-exchange treatment
of the nucleon-nucleon force. Those results were summarised in 
Table~\ref{meson_summary}. 
{}For each nucleus we calculated the rate of change of the binding energy 
with respect to the mass of each meson and the mass of the nucleon itself. 
The values of $K_A$ were obtained by combining the latter with the results 
in Table~\ref{meson_summary}.
\begin{table}[htb]
\begin{center}
\caption{Coefficients $K_{A}$ summarising the rate of variation of 
the binding energies and the $^1$S$_0$ anti-bound state pole 
with respect to quark mass - see Eq.~(\ref{eq:KA}).}
\label{coeff_summary}
\vspace{0.3 cm}
\begin{tabular}{lc}\hline
 Nucleus &~~~ $K_{A}$  \\ \hline \hline
 $D$             & -0.912  \\ \hline
 $T$           &  -0.979 \\ \hline
 $E_P$   & -2.839 \\ \hline
 $^{7}Li$ &    -2.571 \\ \hline
 $^{12}C$ &  -1.438  \\\hline
 $^{16}O$          & -1.082  \\ \hline
\end{tabular}
\end{center}
\end{table}

{}For the deuteron our result, $K_d = -0.91$, is very close to that 
reported by Flambaum and Wiringa~\cite{FW} 
using the AV14 potential, namely 
$-0.84$. Similarly for the triton, our value $K_t=-0.89$ is very close 
to their value of $-0.98$. The closeness of these results for two 
rather different treatments of the NN force lends considerable confidence 
in their reliability. However, for the position of the $^1S_0$ 
anti-bound state 
our calculation differs considerably from that of Ref.~\cite{FW}, taking 
the opposite sign. This suggests that this quantity may be rather more 
model dependent than has been realized hitherto.

In the case of light nuclei, the binding energies reported here were 
calculated in the quark-meson coupling (QMC) model, a relativistic mean-field 
model that takes into account the self-consistent response of the 
internal structure of the nucleon to these mean fields. Through 
the self-consistency, the model yields many-body~\cite{Guichon:2004xg}
or equivalently density-dependent interactions~\cite{Guichon:2006er}. 
Indeed, the density dependent Skyrme forces derived from QMC have proven 
remarkably realistic~\cite{Dutra:2012mb}. The values of $K_A$ deduced 
in this way for $^7$Li, $^{12}$C and $^{16}$O are reported in 
Eqs.~(\ref{ELi7})-~(\ref{EO16}). It is interesting that  
the value obtained for $^7$Li, namely
$K_{^7Li} = -2.57$, is significantly larger than that reported in 
Ref.~\cite{FW}, namely $-1.03$ (AV14) and $-1.50$ (AV18+UIX). 
These authors did suggest that the 
uncertainty on the value of $K$ could be as large as a factor of two 
and our value is consistent at that level. Clearly, this degree of variation 
calls for more investigation to see whether the model dependence can 
be reduced.

\begin{figure}[ht]
\includegraphics[scale=0.7]{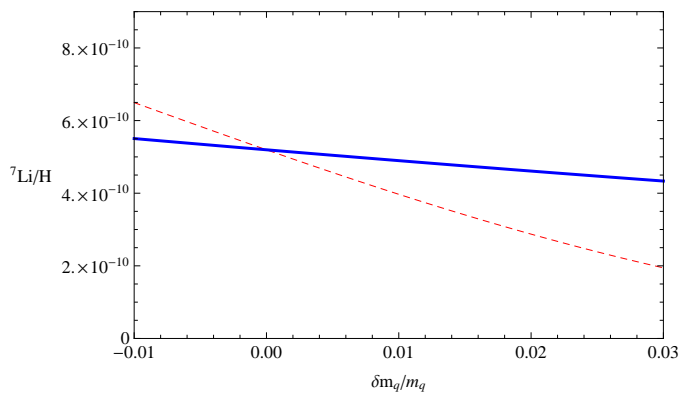}
\caption{(Color online) Abundance of $^{7}$Li with respect to changes 
in the quark mass in $p(n,\gamma)d$ calculated in the same way 
as \cite{Berengut:2009js} (dashed-red line) 
and using our results for $K_{D}$ and $K_{E_P}$ (continuous-blue line).} 
\label{7Li_variation_with_mq}
\end{figure}

Our study of these variations of binding energies with quark mass is, of 
course, motivated by the possible effects on big bang nucleosynthesis (BBN). 
Amongst the many challenges there, the sizeable discrepancy in the abundance of $^7$Li with the latest photon-to-baryon ratio (post WMAP) is of particular 
interest. Figure~\ref{7Li_variation_with_mq} illustrates the $^7$Li 
abundance calculated using the BBN code of Kawano~\cite{Kawano:1992ua}, 
if one allows {\it only} the binding energy of the deuteron  and the 
energy of the virtual $^1S_0$ state to change with 
quark mass. The curves correspond to the values of $K_d$ and 
$K_P$ calculated here (solid line) 
as well as the values used by 
Berengut {\it et al.}~\cite{Berengut:2009js} (dashed line).
The substantial difference in slope means 
that while a 3\% shift in $\delta m_q/m_q$ would suffice to reproduce the 
empirical abundance using  
the values of Berengut {\it et al.}, with our values this would 
require a huge change in quark mass. This simple 
example illustrates the importance of a complete study of the BBN problem 
including all of the consequences of a shift of quark mass within the current 
approach, which we leave for future work. 
Finally, we note that while the variation of the light quark masses should 
be most important, it will also be necessary to take into account the 
effect of a corresponding change in the strange quark mass, especially 
now that the strange quark sigma commutator seems to be under 
control~\cite{Young:2009ps}.

\section*{ACKNOWLEDGEMENTS}
This work was supported by the Australian Research Council through the 
ARC Centre of Excellence in Particle Physics at the Terascale and through
the Australian Laureate Fellowship (FL0992247), as well as by the University 
of Adelaide.

\appendix

\section*{Appendix}
From Table~\ref{tablebind} we find the variations of the binding 
energies for the deuteron and triton ($E_{i}$ with $i=D,T$), and the position of the pole for the $^1$S$_0$ anti-bound state $E_P$; according to changes in the mass of the hadrons ($m_{H}$): 
\be
\frac{\delta E_{i}}{E_{i}} = \frac{1}{E_{i}}\sum_{H} 
\frac{\delta E_{i}}{\delta m_{H}}\delta m_{H} \, .
\ee
We then relate the variation of the mass of each hadron to the variation 
of the quark mass, as given in Eq.~(\ref{vary}):
\be
\frac{\delta m_{H}}{m_{H}} = \nu_{H}\frac{\delta m_{q}}{m_{q}}
\, ,
\ee
so that:
\be
\delta m_{H} = \left(\nu_{H}\cdot m_{H}\right)\frac{\delta m_{q}}{m_{q}} 
\, .
\ee
Combining those results we finally obtain the formula that gives 
rise to the results in Eqs.~\eqref{eq:D_total}, ~\eqref{eq:T_total}  
and ~\eqref{eq:Ep_total}:

\be
\frac{\delta E_{i}}{E_{i}} = \frac{1}{E_{i}}\sum_{H}\frac{\delta E_{i}}{\delta m_{H}}\left(\nu_{H}\cdot m_{H}\right)\frac{\delta m_{q}}{m_{q}}
\, .
\ee

\end{document}